\documentclass[twocolumn,showpacs,preprintnumbers,amsmath,amssymb,superscriptaddress,aps,prl]{revtex4}

\usepackage{graphicx}

\begin{document}

\title{Bloch oscillations of ultracold atoms: a tool for a metrological determination of
$\mathbf{h/m_{Rb}}$}
\author{R\'emy~Battesti}
\affiliation{Laboratoire Kastler Brossel, Ecole Normale
Sup\'erieure, CNRS-UMR 8552, UPMC, 4 place Jussieu, 75252 Paris
Cedex 05, France }
\author{Pierre~Clad\'e}
\affiliation{Laboratoire Kastler Brossel, Ecole Normale
Sup\'erieure, CNRS-UMR 8552, UPMC, 4 place Jussieu, 75252 Paris
Cedex 05, France }
\author{Sa\"\i da~Guellati-Kh\'elifa}
\affiliation{BNM-INM, Conservatoire National des Arts et
M\'etiers, 292 rue Saint Martin, 75141 Paris Cedex 03, France}
\author{Catherine~Schwob}
\affiliation{Laboratoire Kastler Brossel, Ecole Normale
Sup\'erieure, CNRS-UMR 8552, UPMC, 4 place Jussieu, 75252 Paris
Cedex 05, France }
\author{Beno\^it~Gr\'emaud}
\affiliation{Laboratoire Kastler Brossel, Ecole Normale
Sup\'erieure, CNRS-UMR 8552, UPMC, 4 place Jussieu, 75252 Paris
Cedex 05, France }
\author{Fran\c cois~Nez}
\affiliation{Laboratoire Kastler Brossel, Ecole Normale
Sup\'erieure, CNRS-UMR 8552, UPMC, 4 place Jussieu, 75252 Paris
Cedex 05, France }
\author{Lucile~Julien}
\affiliation{Laboratoire Kastler Brossel, Ecole Normale
Sup\'erieure, CNRS-UMR 8552, UPMC, 4 place Jussieu, 75252 Paris
Cedex 05, France }
\author{Fran\c cois~Biraben}
\affiliation{Laboratoire Kastler Brossel, Ecole Normale
Sup\'erieure, CNRS-UMR 8552, UPMC, 4 place Jussieu, 75252 Paris
Cedex 05, France }

\begin{abstract}
We use Bloch oscillations in a horizontal moving standing wave to
transfer a large number of photon recoils to atoms with a high
efficiency ($99.5\%$ per cycle). By measuring the photon recoil of
$^{87}Rb$, using velocity selective Raman transitions to select a
subrecoil velocity class and to measure the final accelerated
velocity class, we have determined $h/m_{Rb}$ with a relative
precision of 0.4 ppm. To exploit the high momentum transfer
efficiency of our method, we are developing a vertical standing
wave set-up. This will allow us to measure $h/m_{Rb}$ better than
$10^{-8}$ and hence the fine structure constant $\alpha$ with an
uncertainty close to the most accurate value coming from the
($g-2$) determination.
\end{abstract}

\maketitle

In the past 20 years, atom manipulation using laser light has led
to the emergence of many powerful techniques  \cite{Chu}. In
particular, it is now possible to observe and measure elementary
processes between light and atoms, such as a coherent momentum
transfer (absorption and emission of a single photon).
Furthermore, by increasing the interrogation time, laser cooling
leads to an improvement of more than two orders of magnitude in
both stability and accuracy in many fields of high precision
measurements \cite{Santarelli,Oates}. These advances allow us to
measure precisely the recoil velocity $v_r$ of the atom absorbing
or emitting a photon (~$v_r = \hbar k /m$, where $k$ is the wave
vector of the photon absorbed by the atom of mass $m$). Such a
measurement yields a determination of $h/m$ which can be used to
infer a value for the fine structure constant $\alpha$ via
\cite{Taylor}:
\begin{equation}
       \alpha^2=\frac{2R_\infty}{c}\frac{M}{M_e}\frac{h}{m}
\label{eqn1}
\end{equation}

M and m are respectively, the mass of test particle in atomic and
SI Units. In this expression several terms are known with a very
small uncertainty:  $8 \times 10^{-12}$ for the Rydberg constant
$R_\infty$ \cite{Schwob, Udem} and $7 \times 10^{-10}$ for the
electron mass $M_e$ \cite{Beier,Farnham}. A recent measurement
using Penning trap single ion spectrometry allows a determination
of $M_{\text{Rb}}$ with an uncertainty less than $2 \times
10^{-10}$ \cite{Bradley}. In short, the determination of $\alpha$
using this formula is now limited by the uncertainty in the ratio
$h/m$ \cite{Taylor}.

The fine structure constant can be deduced from experiments
related to different branches of physics (QED, Solid state
physics,...) \cite{Liu,Kinoshita,VanDick,Kruger,Shields,Jeffery}.
Many of these measurements lead to determinations of $\alpha$ with
a relative uncertainty on the order of $10^{-8}$ but their total
dispersion exceeds $10^{-7}$(CODATA 98, \cite{Mohr}). In order to
test the validity of these different measurements, a new accurate
determination of $\alpha$ is highly desirable.

The recoil of an atom when it absorbs a photon was first observed
in the recoil-induced spectral doubling of the $CH_4$ saturated
absorption peaks \cite{Hall}. Since then, almost all recent
measurements of the recoil velocity have been based on atomic
interferometry \cite{Borde} using stimulated Raman transitions
between two hyperfine ground state levels \cite{Young,Wicht}. The
precision of these experiments is increased by giving additional
photon recoils to the different interferometer paths. In this
process, the efficiency of the recoil transfer is a crucial
parameter.
 Recently, S.~Chu group at Stanford, using a coherent adiabatic
 transfer technique and high intensity Raman transitions
 pulses, has achieved an efficiency of $94\%$, allowing
 a total momentum transfer of 120 recoils, and hence
 an absolute accuracy of $7.4$ parts per billion in $\alpha$ \cite{Wicht}.
D. Pritchard and colleagues have developed another tool for a
determination of $h/m_{Na}$, using a Bose-Einstein condensate as a
bright subrecoil atom source in the "contrast interferometry"
technique \cite{Gupta}. This experiment seems, presently, to be
limited by the low momentum
 transfers.

In this paper, we investigate the phenomena of Bloch oscillations
of atoms driven by a constant inertial force in a periodic optical
potential \cite{Ben Dahan}. This method is based on stimulated
Raman transitions, induced by counterpropagating laser beams,
involving only one hyperfine level in order to modify the atomic
momentum, thus leaving the internal state unchanged. The atoms are
coherently accelerated using a frequency-chirped standing wave. In
order to compensate the Doppler effect, the frequency difference
between the two beams is increased linearly. Consequently, the
atoms are resonant with the beams periodically. This leads to a
succession of rapid adiabatic passages between momentum states
differing by $2\hbar k$ ($2v_r$ in terms of velocity). As
explained above, the final accuracy is determined by the number of
additional recoils, which strongly depends on the efficiency
population transfer between momentum states. In our experiment, we
achieve an efficiency of $99.5\%$ per Bloch oscillation; which may
provide a great opportunity for high precision measurement of the
recoil shift.

\begin{figure}
\includegraphics[width=1\columnwidth]{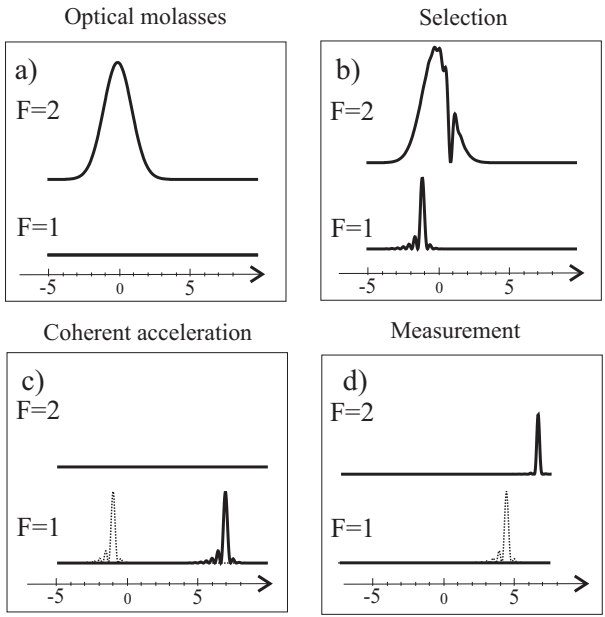}
\caption{\label{SeqTemp} Evolution of the velocity distribution
(in $v_r$ unit) during the experiment: a) Initial velocity
distribution, the atoms are in $5^2\text{S}_{1/2}$, $\left|F=2,
m_F = 0\right>$ state, b) Subrecoil selection, the atoms are
transferred from $5^2\text{S}_{1/2}$, $\left|F=2, m_F = 0\right>$
to $5^2\text{S}_{1/2}$, $\left|F=1, m_F = 0\right>$, c) Coherent
acceleration for $N=4$ Bloch oscillations, the atoms are in
$5^2\text{S}_{1/2}$, $\left|F=1, m_F = 0\right>$ and d)
Measurement of the final velocity class, the atoms are transferred
from $5^2\text{S}_{1/2}$, $\left|F=1, m_F = 0\right>$ to
$5^2\text{S}_{1/2}$, $\left|F=2, m_F = 0\right>$}
\end{figure}

The details of the experiment have been described previously
\cite{Battesti}. The set-up for the laser cooling uses a
magneto-optical trap (MOT) in a Rubidium vapor cell. After a few
seconds, the MOT is loaded and then the magnetic field is switched
off, leaving the atoms to equilibrate in an optical molasses, at a
temperature of about $3~\mu \text{K}$. The sequence then involves
three steps (see Fig.\ref{SeqTemp}): first we select a narrow
subrecoil velocity class with a well defined mean initial velocity
using a Raman velocity-selective $\pi$-pulse \cite{Kasevich}. Next
we transfer $2 N$ photon recoils by a coherent acceleration of
atoms (Bloch oscillations). Finally, we measure the final velocity
class, using another Raman velocity-selective $\pi$-pulse.

For the initial selection and the final measurement, the two Raman
beams are generated by two Master Oscillatory Power Amplifiers
(MOPAs) injected by two grating-stabilized extended-cavity laser
diodes (ECLs). One of the two diodes is frequency stabilized on a
highly stable Zerodur Fabry-Perot Cavity (ZFPC). This cavity was
calibrated using different optical references, allowing a
determination of the laser frequency with an accuracy better than
10~MHz ($3 \times 10^{-8}$). A heterodyne signal and a frequency
chain around the Rubidium hyperfine splitting (6.8~GHz) are used
to phase lock the second ECL to the first one. All auxiliary
sources in the frequency chain are referenced to the same stable
10~MHz quartz oscillator. To reduce spontaneous emission and light
shift, the ECLs are detuned by about 340~GHz from the D2 line. The
MOPA beams are sent through two 80~MHz acousto-optic modulators
for timing and intensity control. Their radio frequencies are also
referenced to the 10~MHz quartz oscillator. The two beams are
coupled in an optical fiber, they have linear orthogonal
polarizations and their intensities are actively stabilized.

 In order to perform the selection phase, we use a
square Raman pulse with a frequency initially fixed at
$\delta_{\text{select}}$. For a detuning of 340~GHz and an
intensity of 120~mW/cm$^{2}$, the $\pi$ condition is achieved
using a $T=1.7$~ms pulse. Such a pulse transfers atoms from state
$5^2\text{S}_{1/2}$, $\left|F=2, m_F = 0\right>$ to
$5^2\text{S}_{1/2}$, $\left|F=1,m_F = 0\right>$, with a velocity
dispersion of about $v_r/30$ centered around $(\lambda
\delta_{\text{select}}/ {2}) - v_r$ where $\lambda$ is the laser
wavelength. In this horizontal geometry, the width of the
transferred velocity class, which is proportional to $1/T$, is
only limited by the fall of the atoms through the lasers beams.
The value of $T$ represents a good compromise between resolution
(the width of the selected velocity distribution) and the
signal-to-noise ratio (proportional to the number of selected
atoms).

After the Raman selection process, a beam resonant with
$5^2\text{S}_{1/2}$, $F=2$ to $5^2\text{P}_{3/2}$, $F=3$ cycling
transition pushes away atoms remaining in state $F=2$. The
selected atoms are then exposed to two counterpropagating beams
generated by a Ti-Sapphire laser whose frequency is also
stabilized to the ZFPC. This laser beam is split in two, each beam
passing through an acousto-optic modulator to control its
frequency. In order to perform a coherent acceleration, we vary
the frequency difference between the two Bloch beams linearly with
time: $\Delta\nu (t)= 2 a t/\lambda$ where $a$ is the effective
acceleration. The two beams are superimposed onto the horizontal
optical axis of the selection Raman beams using the same optical
fibers. The two Bloch beams have the same linear polarization,
equal intensity (160 mW for each beam) and are red detuned by 100
GHz from the $5^2\text{S}_{1/2}$ to $5^2\text{P}_{3/2}$ resonance
line. The duration of the acceleration process is typically 4.4
ms. The optical potential is adiabatically turned on in about
$300~\mu s$.

In the case where the constant inertial force seen by the atoms is
weak enough, all the selected atoms are accelerated. In a Bloch
oscillation scheme, this is equivalent to avoiding interband
transitions. This condition may be expressed in the weak binding
limit \cite{Peik} by:
\begin{equation}
       \pi\frac{d\Delta\nu(t)}{dt}<< \left(\frac{U_0}{2\hbar}\right)^2
\end{equation}
where $U_0$ is the depth of the potential induced by the light
shift due to the standing wave. In this limit, the interband
transition rate per Bloch period is given by a Landau-Zener
formula
 $R=e^{(-a_c/a)}$ where $a_c$ is the critical
acceleration, proportional to $(U_0/E_r)^2$ ($E_r$ is the recoil
energy) \cite{Ben Dahan}. In our experiment $U_0$ is about
$11~E_r$  and atoms acceleration is $133$~ms$^{-2}$ leading to a
theoretical efficiency of $99.9\%$ per oscillation.

\begin{figure}
\includegraphics[width=1\columnwidth]{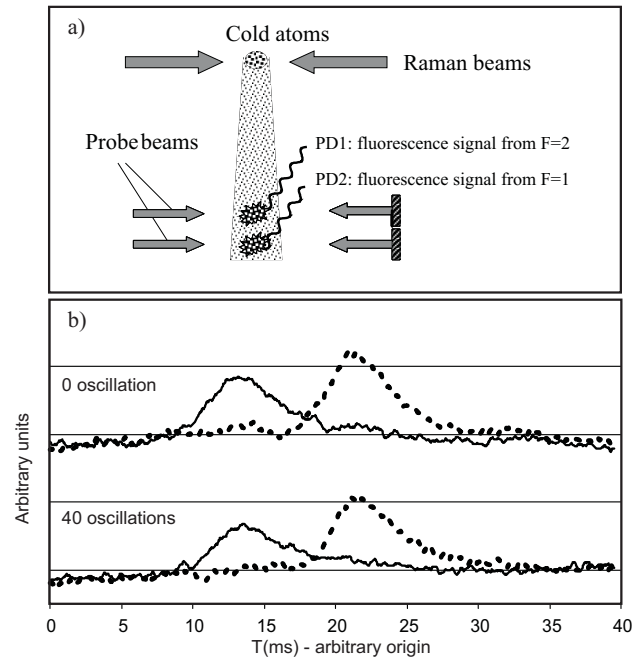}
\caption{\label{tofsExp} a) Experimental scheme. b) Time of flight
signal for $N = 0$ and $N = 40$ Bloch oscillations. The dashed
line corresponds to the signal of the $F=1$ atoms and the solid
line for the $F=2$ atoms which are transferred from the hyperfine
state $F=1$ by the second Raman pulse. After 40 Bloch oscillations
the time of flight signal remains almost unchanged, emphasizing
thus the high efficiency transfer of our experiment. The analysis
of these signals gives the number of atoms in each hyperfine
state.}
\end{figure}

After the acceleration process we perform the final velocity
measurement using another Raman $\pi$-pulse, whose frequency is
$\delta_{\text{measure}}$. Population transfer from the hyperfine
state F=1 to the hyperfine state F=2 due to the second Raman pulse
is maximal when
$2\pi(\delta_{\text{select}}-\delta_{\text{measure}}) = 2N\times
(\mathbf{k_1}-\mathbf{k_2}) \cdot\mathbf{k_{Bloch}} \hbar/m_{Rb}$,
where $\mathbf{k_1}$, $\mathbf{k_2}$ and $\mathbf{k_{Bloch}}$ are
respectively the wave vectors of the Raman and Bloch beams. The
populations ($F=1$ and $F=2$) are measured separately by using the
one-dimensional time of flight technique developed for atomic
clocks and depicted in \cite{Clairon}. The detection zone is 15 cm
below the center of the trap (Figure \ref{tofsExp}). To avoid the
horizontal motion of the atoms, and in order for the atoms to
reach the detection zone, a symmetric acceleration-deceleration
scheme is used~: instead of selecting atoms at rest, we first
accelerate them to $2 N\,v_r$, using N Bloch oscillations, then we
make the three steps sequence: selection, coherent deceleration
($N$ Bloch oscillations) and measurement, according to
Fig.~\ref{SeqTemp}.

Fig.~\ref{tofsExp}b shows a typical time of flight signal for
$F=2$ (left peak) and $F = 1$ (right peak) when the second Raman
frequency is centered at the top of the final velocity
distribution. In this figure, we present the signal for $N=0$ and
for $N = 40$ Bloch oscillations. Comparing the number of atoms
between the two situations, we demonstrate that the losses during
the Bloch acceleration for $N = 40$ are less than $20\%$,
corresponding to a transfer efficiency of about $99.5\%$ per
oscillation. These losses are not due to the spontaneous emission,
which is evaluated to $0.1\%$ per oscillation, but probably to the
residual horizontal displacement of the atoms (about 5 mm) after
the acceleration-deceleration process.

\begin{figure}
\includegraphics[width=1\columnwidth]{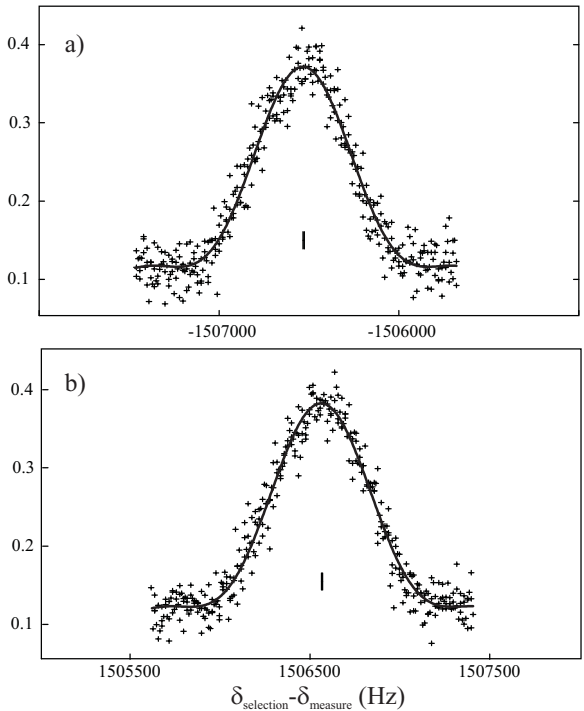}
\caption{\label{spec50bloch} Final velocity distribution for 50
Bloch oscillations, in both directions: a)Around $2 N v_r$ and
b)Around $- 2 N v_r$. The center of the final velocity
distribution can be located with an uncertainty of 3~Hz. The
photon recoil frequency deduced from these two measurements is
15066.690(23)~Hz. }
\end{figure}

To reduce systematic errors, we perform an alternate and symmetric
recoil transfer in both horizontal opposite directions. We
determine the recoil frequency by a differential measurement of
the center of the two final velocity distributions.
Fig.\ref{spec50bloch} shows a typical scan of final velocity
distribution for $N=50$ Bloch oscillations for both directions.
Each of the 400 data points corresponds to a single cycle
(cooling, selection, acceleration of $2N v_r$, and measure). From
a data analysis of 200 points (10 minutes) we can split the final
velocity distribution with an uncertainty of $v_r/5000$. Hence,
the relative uncertainty of the measurement of $v_r$ is $1.5
\times 10^{-6}$.

\begin{figure}
\includegraphics[width=\columnwidth]{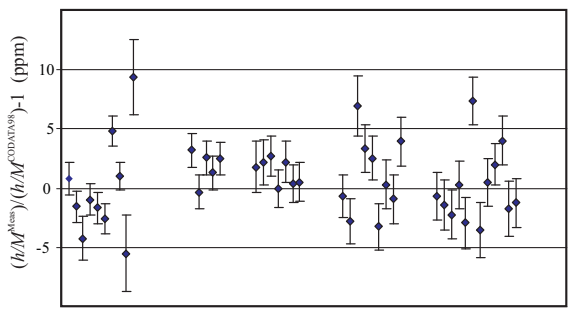}
\caption{\label{Rb87mars} Chronological display of the data taken
for $N=50$ Bloch oscillations in both directions. The mean
deviation from the expected value is $6.1(4.2) \times 10^{-7}$,
with $\chi^2=99$.}
\end{figure}

Fig.~\ref{Rb87mars} shows, chronologically, 43 determinations of
$h/m_{Rb}$ using such measurements, compared to the expected value
of $h/m_{Rb}$, using the CODATA 98 value of $\alpha$. The mean
value lies $6.1\times 10^{-7}$ above the expected value with a
relative uncertainty of $4.2\times 10^{-7}$.
We have estimated errors form wave front curvature (1.3~ppb),
differential light shift (80~ppb), Zeeman effect (57~ppb) and
laser frequencies (52~ppb). They are an order of magnitude lower
compared to this disagreement. We believe that this disagreement
and the dispersion of the results ($\chi^2=99$ for $43$
measurements) can be explained by
 the systematic variations of the initial velocity
distribution of the cold cloud and the phase fluctuation of the
Raman beams. These effects have not yet been quantified.

In conclusion, we have demonstrated that coherent acceleration
using Bloch oscillations is a powerful method to transfer a large
number of additional photon recoils to atoms. In the horizontal
scheme, the number of momentum recoil transfers is limited by the
fall of atoms. In order to take advantage of the high transfer
efficiency of the Bloch oscillation technique, we plan to build a
set-up with vertical Bloch and Raman beams. In this case, the
number of additional recoils will be limited by the transverse
motion, and we can increase the $\pi$-pulse duration in order to
select a narrower velocity class. Moreover, due to the
gravitational acceleration $g$, the vertical motion is more
complicated and this geometry provides scope for two different
experiments: either the atoms are accelerated by a moving standing
wave as in the horizontal scheme, or they are placed in a pure
standing wave. In this case, the atoms oscillate around the same
position at the frequency $mg/2\hbar k$ \cite{Battesti}.
Furthermore, there is no significant displacement of the atom
between the velocity selection and measurement, and thus, several
systematic effects are reduced. On the other hand, a precise
determination of the local gravity field is required to exploit
this technique to the full. Finally we expect to increase the
number of transferred recoils up to $N=500$, to obtain a
determination of $v_r$ with an uncertainty better than $10^{-8}$,
leading to a determination of $\alpha$ with an uncertainty of
about 5~ppb, close to the more accurate value deduced from ($g-2$)
\cite{Kinoshita,VanDick,Mohr}

\begin{acknowledgments}
We thank A.~Clairon, C.~Salomon and J.~Dalibard for valuable
discussions. This experiment is supported in part by the Bureau
National de M\'etrologie (contract 993009) and by the R\'egion Ile
de France (contract SESAME E1220).
\end{acknowledgments}

\end{document}